\documentclass[useAMS,usenatbib,usegraphicx]{mn2e}
\usepackage{color}
\title[3.3$\,\mu$m PAH emission of the Red Rectangle]{Spatial distribution and interpretation of the 3.3$\,\umu$m PAH emission band of the Red Rectangle}

\author[A. Candian, T.H. Kerr, I.-O. Song, J. McCombie and P.J. Sarre]{A. Candian$^{1}$\thanks{alessandra.candian@gmail.com}, T.H. Kerr$^{2}$, I.-O. Song$^{1}$\thanks{current address: Korea Science Academy of KAIST, 899 Danggam 3-dong, Busanjin-gu, Busan 614-822, Korea}, J. McCombie$^{1}$ and P.J. Sarre$^{1}$\thanks{Peter.Sarre@nottingham.ac.uk}\\
$^{1}$School of Chemistry, The University of Nottingham, University Park, Nottingham NG7 2RD, U.K.\\
$^{2}$United Kingdom Infrared Telescope, Joint Astronomy Centre, 660
N. A'ohoku Place, University Park, Hilo, Hawaii 96720, U.S.A.\\
}

\begin{document}

\date{Accepted 2012 July 18.  Received 2012 July 18; In original form 2011 September 1 }

\pagerange{\pageref{firstpage}--\pageref{lastpage}} \pubyear{2012}

\maketitle

\label{firstpage}

\begin{abstract}
The spatial distribution of 3.3$\,\umu$m PAH and associated emission in the $3.3''\times\,6.0''$ inner region of the Red Rectangle nebula has been determined using the UIST imager-spectrometer at the United Kingdom Infrared Telescope (UKIRT). Interpretation of the 3.3$\,\umu$m feature as comprising two spectroscopic components centred at 3.30$\,\umu$m and 3.28$\,\umu$m, as put forward by Song et al. (2003, 2007), is supported by these data which reveal that they have different spatial distributions. It is deduced that there are two classes of 3.3$\,\umu$m band carrier with a peak wavelength separation of $\sim$0.02$\,\umu$m. From comparison of the 3.3$\,\umu$m observations with laboratory and theoretical spectra for a range of PAH molecules it is proposed that the 3.28$\,\umu$m and 3.30$\,\umu$m components arise from `bay' and `non-bay' hydrogen sites, respectively, on the periphery of small neutral PAHs. Observational data are also obtained for \emph{L}-band continuum emission and for the Pfund$\,\varepsilon$ hydrogen recombination line.

\end{abstract}


\section{Introduction}

The 3.3$\,\umu$m emission band is a well-known member of the family of unidentified infrared `UIR' or `Aromatic IR' bands (AIBs) that are attributed to polycyclic aromatic hydrocarbon (PAH) molecules.  It is the shortest wavelength UIR band and is assigned to the infrared active C--H stretching modes. This feature, in common with other major UIR bands at 6.2, 7.7, 8.6 and 11.2$\,\umu$m, is ascribed to PAHs at a generic rather than specific molecular level - for discussion see: \citet{tok97,tie08,wal09,pee11}.  Only in the cases of benzene (C$_6$H$_6$) in CRL~618 \citep{cer01} and SMP LMC 11 \citep{bern06}, C$_{60}$ and C$_{70}$ in the planetary nebula Tc~1 \citep{cam10} and C$_{60}$ in the reflection nebulae NGC~7023 and NGC~2023 \citep{sel10} and in other objects, have specific C$_6$~-~ring molecules been identified.

The biconical Red Rectangle nebula, centred on the binary star HD~44179, is a remarkable object with a very intense display of UIR bands. Long-slit studies of the 3.3$\,\umu$m emission feature of the Red Rectangle and other objects \citep{son03,son07}, have suggested that the band profile can be interpreted in terms of two emission components centred at 3.28$\,\umu$m and 3.30$\,\umu$m\footnote[1]{As in \citet{son07} we use the nominal description `3.3' to refer to the whole 3.3$\,\umu$m profile, with the wavelengths of the two components derived from profile fitting written with a second decimal place \emph{viz} 3.28 and 3.30$\,\umu$m.}. Exposures with a slit aligned along the NW whisker of the nebula revealed that 3.30$\,\umu$m emission appears on-star and that the 3.3$\,\umu$m feature changes shape with offset due to addition of a 3.28$\,\umu$m component. Although this two-component description works well as a basis for interpreting the spectra, it does not preclude weak emission at slightly different wavelengths and there are indications of some wavelength variation as a function of offset \citep{son07}.

 In this Integral Field Unit (IFU)-based study the spatial distribution of the 3.3$\,\umu$m band is investigated over the inner part of the nebula ($3.3''\times6.0''$), centred on the star HD~44179.  The aim is to improve knowledge of the distribution of PAHs in the Red Rectangle and of the size, shape and charge state of 3.3$\,\umu$m band carriers both in this object and more generally in Galactic and extragalactic sources.  The paper is arranged first with a review of previous observational, experimental, theoretical and modelling studies of the 3.3$\,\umu$m band (Section 2), followed by observational details (Section 3) and the derived IFU images and their spectral and spatial decomposition (Section 4).  In Section 5 we consider the origin of the 3.3$\,\umu$m band with particular reference to experimental gas-phase spectroscopic data for small PAHs and their bay/non-bay character, followed by DFT calculations of vibrational frequencies and transition strengths and the incorporation of this experimental and theoretical information into a UV-pumped emission model (Section 6). Comparison is made between the 3.3$\,\umu$m astronomical images and those for `blue luminescence' (Section 7), and a summary of results is given in Section 8.

\section{Overview}

In order to provide a backdrop for the following Sections the current state of knowledge of the 3.3$\,\umu$m PAH feature is summarized here from observational, laboratory, theoretical and modelling perspectives.

\subsection{Astronomical studies of the 3.3$\,\umu$m band}

Using ground-based studies \citet{tok91} made the first classification of the 3.3$\,\umu$m band and identified a common Type~1 and a rare Type~2. Following \emph{ISO} observations, an (\textbf{A, B}) classification scheme based on the band shape and central wavelength was put forward for a large number of objects by \citet{van04}, with designations \textbf{A$_{3.3}$}, \textbf{B1$_{3.3}$} and \textbf{B2$_{3.3}$}. The \textbf{A$_{3.3}$} set is common, has an approximately symmetrical profile with peak position of $\sim\,3.290\,\umu$m and a FWHM of 0.040$\,\umu$m; it is similar to Type~1. In their two-component interpretation, \citet{son07} showed that this corresponds to approximately equal contributions of 3.28$\,\umu$m and 3.30$\,\umu$m `carriers', which results in a peak wavelength of $\sim\,3.29\,\umu$m.

A smaller group of objects was classed as \textbf{B1$_{3.3}$} with maximum intensity at slightly longer wavelength ($\sim\,3.293\,\umu$m) and with a somewhat narrower width than for \textbf{A$_{3.3}$}.  In the Song et al. (2007) interpretation this corresponds to a smaller relative contribution from the 3.28$\,\umu$m component.  The class \textbf{B2$_{3.3}$} has only one member and is found for \emph{ISO} spectra of the Red Rectangle.  However, these data comprise a $14''\times\,20''$ exposure of both star and nebula and the profile in fact varies within the object \citep{son03}; the peak wavelength is yet longer at $\sim\,3.297\,\umu$m (FWHM $\sim0.037\,\umu$m) and lies fairly close to a `pure' 3.30$\,\umu$m component as found for on-star PAH emission by \citet{son03}. It is similar to the rarer Type~2 of \citet{tok91}. The overall 3.3$\,\umu$m profile in the Red Rectangle develops from Type~2 ($\sim\,$\textbf{B2$_{3.3}$}) on-star towards Type~1 (or approximately) \textbf{B1$_{3.3}$} at $\sim\,7.5''$ offset.  Expressed in terms of spectral components, for the Red Rectangle the Lorentzian-shaped 3.30$\,\umu$m feature dominates on-star and the 3.28$\,\umu$m feature grows (relatively) with offset, reaching about 50\% of the 3.30$\,\umu$m intensity at $\sim\,7.5''$.

The general sequence \textbf{A$_{3.3}$~--~B1$_{3.3}$~--~B2$_{3.3}$} can be summarized semi-quantitatively in terms of a 3.28/3.30 intensity ratio as $\geq1$, $\sim0.5$ and $\leq0.2$, respectively, although a continuous range of ratios is very likely to exist among different astronomical sources, and even within a single object as occurs in the Red Rectangle.

\subsection{Laboratory studies of the 3.3$\,\umu$m band}

Numerous experiments have been conducted in absorption on individual PAH molecules trapped in cold inert gas matrices, \emph{e.g. } neutral PAHs with 2-4 rings \citep{hud98a} and $\geq$~5~rings \citep{hud98b}. Although somewhat removed from gas-phase PAHs under interstellar conditions because of the extremely low temperature and matrix perturbations, these measurements allow easy comparison of infrared spectra of a wide range of both neutral and ionised PAHs. The most extensive experimental information for the 3.3$\,\umu$m band of PAHs is given in the NASA Ames Infrared Spectroscopic Database \citep{bau10}.

Experiments have been conducted on the 3.3$\,\umu$m vibrational band of gas-phase PAHs in both absorption \emph{e.g.} \cite{sem91,job94,job95,can96,can97,hun04,pir09,ste11} and in emission \emph{e.g.} \cite{sch94,wil95,coo96,coo98a,wag00,kim02}.  In a close simulation of interstellar conditions, UV laser excitation of cold gas-phase naphthalene in a supersonic jet \citep{sch94} yielded a single-peaked IR emission band centred at 3.30$\,\umu$m (see their Figure~1~A); notably it is significantly red-shifted with respect to the gas-phase absorption spectrum at $\sim$500~K which peaks at 3.26$\,\umu$m \citep{ste11} and a jet-cold gas-phase spectrum which shows two resolved peaks at 3.26 and 3.25$\,\umu$m.  As shown by \citet{job95}, it is not expected that gas-phase C--H absorption wavelengths at low temperatures will match astronomical observations as there is a marked dependence of the peak wavelength on temperature. Given that the astronomical infrared emission is necessarily from significantly heated PAH molecules following UV/visible photoexcitation, in this work we have chosen to compare astronomical spectra with those of heated PAHs, where available, and this is discussed further in Section 5.

Experimental work on condensed-phase carbon material in both absorption and emission reveals a C--H stretching feature which matches quite well the astrophysical data \citep{sak90,got97,sco97,hu08,dul11}.  However, the physical state of the material is significantly different from the gas-phase PAHs which are more commonly invoked as UIR band carriers.

\subsection{Theoretical studies of the 3.3$\,\umu$m band}

Although high-level \emph{ab-initio} calculations of C--H vibrational frequencies and  intensities have been undertaken \citep{def93,pau97,pau11}, these are necessarily limited to small systems.  The overwhelming majority of computations have been made using density functional theory (DFT) mostly with B3LYP/4-31G  \citep{lan96,bau00,bau02}. The most extensive theoretical information for vibrational transitions for PAHs is given in the NASA Ames Infrared Spectroscopic Database \citep{bau10}. A complementary database by \citet{mal07} includes energetic, rotational, vibrational and electronic properties of PAHs in a wide range of charge states.

In a recent study, vibrational frequencies for large `regular' and `irregular' PAHs have been calculated for PAHs with up to 130 carbon atoms including the 3.3$\,\umu$m band \citep{bau08,bau09}. These authors conclude that cationic, anionic and neutral PAHs all contribute to the 3.3$\,\umu$m feature, the anion to neutral ratio being estimated to be $\sim0.7$.  It was suggested that the relative contributions of anionic and neutral forms for compact PAHs may be responsible for the Type 1 and Type 2 band profiles of \cite{tok91}, where the shorter wavelength contribution would be due to neutrals. It was also concluded that `the vast majority of astronomical PAHs responsible for the emission in the 3$\,\umu$m region are compact, not irregular, elongated, or bent'. This deduction was based on the relative flux  of the continuum-subtracted 3.3$\,\umu$m feature to the residual above the continuum at 3.23 and 3.24$\,\umu$m, respectively, for neutrals and anions, and where a 15~cm$^{-1}$ red shift was invoked.

\subsection{Modelling studies of 3.3$\,\umu$m emission}

A number of studies have been directed towards understanding astronomical 3.3$\,\umu$m emission, usually as one of a wider range of UIR bands - see for example \citet{coo98b,pec02,pat08,bau10}. These studies show how the 3.3$\,\umu$m band is dominated by small PAHs, which can reach peak temperatures high enough to allow emission in high frequency modes \citep{sch93}. The study of most relevance to this work is that of \citet{mul06} in which infrared emission fluxes for anthracene, phenanthrene and pyrene were computed in detail with reference to the radiation field in the Red Rectangle.  Stimulated in significant part by the discovery of `Blue Luminescence' (BL) and the suggestion that this is due to emission from molecules with three or four aromatic units \citep{vij04,vij05}, \citet{mul06} showed that the computed 3.3$\,\umu$m fluxes for anthracene, phenanthrene and pyrene were reasonably consistent with \emph{ISO} data for 3.3$\,\umu$m emission of the Red Rectangle. The possible link between BL and the 3.3$\,\umu$m PAH band is discussed further in Section 7.

\section[]{IFU Observational details}

HD~44179 was observed on three separate occasions, 2005 February 19, February 22 and March 5, using the UIST imager-spectrometer at the 3.8~m United
Kingdom Infrared Telescope (UKIRT). The observations were carried out using the integral field unit (IFU) in combination with the short-\emph{L} grism, giving
a central wavelength of 3.27$\,\umu$m and a wavelength coverage of approximately 0.7$\,\umu$m. The UIST IFU covers a rotatable 3.3$''\times\,6.0''$ region
of sky with a plate scale of 0.12$''\times\,$0.24$''$. Light incident on the IFU is then fed to the 14 useable 0.24$''\times\,6.0''$ slitlets, then onto the
dispersing grism (R $\sim1400$) and finally onto a $1024\times1024$ pixel InSb array. A spectral image, individual spectra and a ``white-light'' image can
thus be derived from the data.

HR~2491 (A1V, \emph{K}~=~-1.39) and HR~1713 (B8~Iab, \emph{K}~=~0.213) were observed at matching airmass both before and after the observations of HD~44179 in order to remove telluric features and for flux calibration. The image rotator was used to align the long-axis of the IFU with the outflow from HD~44179
(position angle 13$^{\circ}$ E of N) and the telescope was nodded approximately 7$'$~NW for sky frames due to HD~44179's extended nature. An ambiguity was noted in the recorded coordinate system which affected the N--S orientation.  Fortunately this was completely resolved as the emission flux in the Southernmost lobe in two earlier IR imaging studies has been shown to be about twice as high as for the Northern equivalent \citep{mek98,men98}. On the first two nights $10\,\times\,$2-sec exposures were taken at each nod position which was later optimised to 2$\,\times\,$8-secs for the final night giving a total on-source exposure time of 4544~s. An internal halogen black-body source was used to flatfield the data and wavelength calibration was achieved \emph{via} identification of sky lines. Data were reduced initially using UKIRT's ORAC-DR data reduction pipeline and then further reduced and analysed using {\sc figaro}, {\sc kappa} and {\sc datacube} routines in the STARLINK software distribution.The \emph{L}-band seeing for 19 and 22 February was approximately 0.7$''$ and sub-0.5$''$ for 5 March 2010.  For ease of discussion, in the following text the description `N' and `S' is used even though the orientation of the biconical axis is tilted 13$^{\circ}$ with respect to the actual N--S axis.

\begin{table}
\caption{Wavelength ranges selected for the spectral images.}
\begin{center}
\begin{tabular}{lcp{2.0cm}}
 \hline
 Feature           &  Origin       &  Range ($\umu$m) \\
 \hline
 L-band continuum  &  Dust       &             2.880-3.240; 3.540-3.610 \\
\\
  3.3$\,\umu$m (total)    &   C-H stretch     &     3.260-3.350   \\
\\
  3.28 \& 3.30$\,\umu$m     &   C-H stretch     &  Deconvolution (see text)\\
\\
  3.4$\,\umu$m      &  C-H$_{2,3}$ stretch     &     3.375-3.427 \\
\\
  Pfund$\,\varepsilon$ & H recombination &  2.980-3.100  \\
\hline
\end{tabular}
\end{center}
\label{tab:tab1_label}
\end{table}

Table~\ref{tab:tab1_label} gives the wavelength ranges used in producing spectral maps for the continuum and the PAH bands at 3.3$\,\umu$m. The spectral maps for the `3.28' and `3.30'$\,\umu$m components were produced by performing a robust double-Lorentzian fit to the full 3.3$\,\umu$m band using the IDL routine {\sc mpfitexpr} \citep{mark09}. The 3.4~$\mu$m band which appears in spectra offset from the central star was also considered. The flux-calibrated spectrum produced by the pipeline showed the presence of the Pfund$\,\varepsilon$ hydrogen emission line at 3.04~$\mu$m, for which a continuum-subtracted image was also produced.

\section{IFU Imaging Results}

\begin{figure}
\includegraphics[scale=0.4]{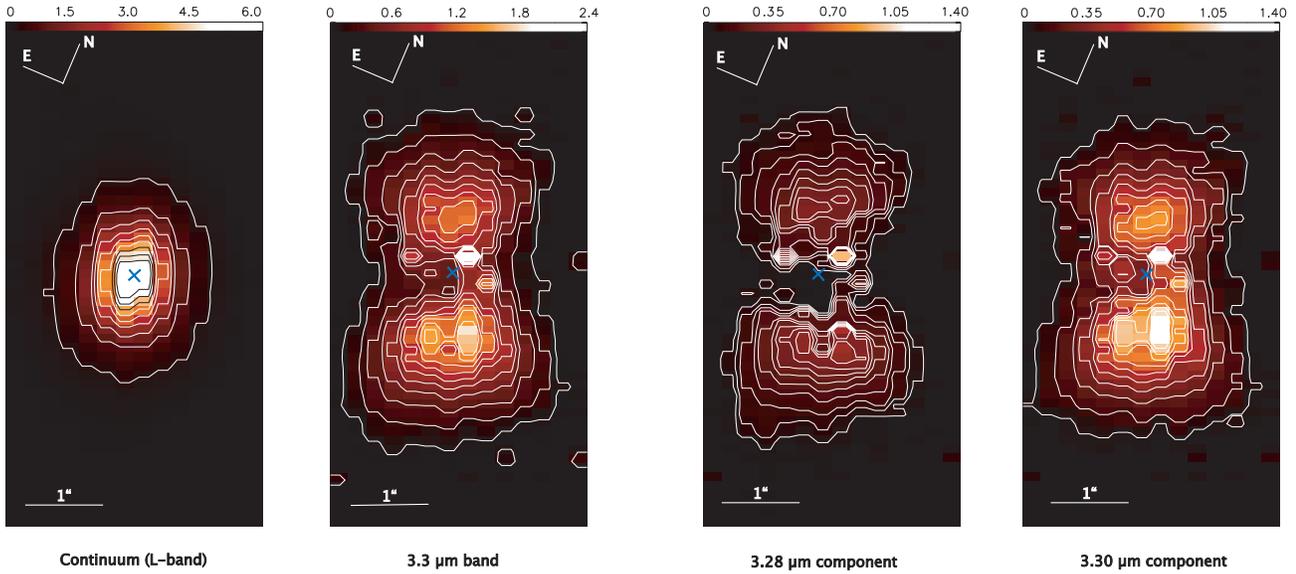}
\caption{Left panel. Spectral map ($3.3''\times\,6.0''$) of the continuum emission, integrated in the \emph{L}-band. The colour scale is in units of 10$^{-13}~$Wm$^{-2}\umu$m$^{-1}$. The contours represent levels from 5 to 95\% of the maximum flux, with steps of 10\%. The intensity maximum is centred on HD~44179 (blue cross) and extends gradually in the outer region as an ellipse with major axis 1.35$''$ and minor axis 1.1$''$. Right panel. Spectral map of the 3.3$\,\umu$m band, with colour scale in units of 10$^{-11}~$Wm$^{-2}\umu$m$^{-1}$.  The contours are as in the left panel. The contours form a bipolar shape, with the Southern lobe somewhat brighter. Colour versions of all figures are available in the on-line paper.}
\label{fig1}
\end{figure}

\begin{figure}
\includegraphics[scale=0.4]{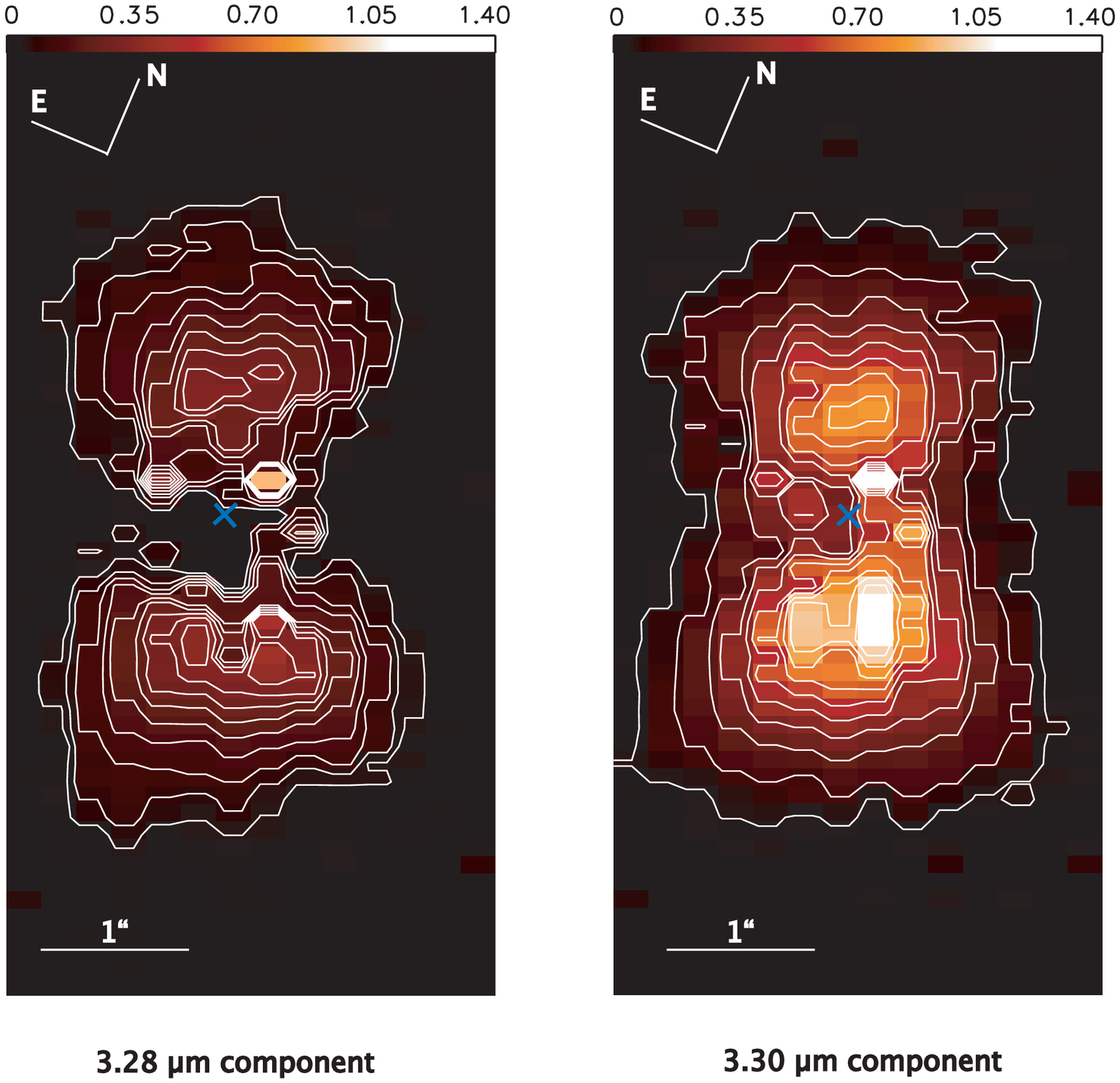}
\caption{Left panel. Spectral map (3.3$''\times\,$6.0$''$) of the 3.28~$\umu$m component of the 3.3$\,\umu$m emission of the Red Rectangle. The contours represent levels from 15\% to 95\% with steps of 10\%.  Two bright peaks (separated by 1.3$''$) appear north and south of the central star, but there is no emission from the centre. Right panel. Spectral map of the 3.30$\,\umu$m component; the contours correspond to levels from 5\% to 95\% in steps of 10\%. The image is similar to that of the 3.3$\,\umu$m band. The colour scale is in units of 10$^{-11}~$Wm$^{-2}\umu$m$^{-1}$ for both maps.}
\label{fig2}
\end{figure}

\begin{figure}
\includegraphics[scale=0.35]{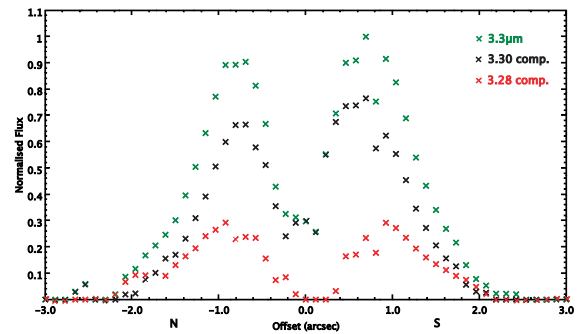}
\caption{Normalised intensity profiles derived from the spectral maps, cut along the major axis at P.A.~=~13$^{\circ}$. Comparison between the N-S profiles for the 3.3$\,\umu$m band, and the 3.30 and 3.28$\,\umu$m components is shown.}
\label{fig3}
\end{figure}

\subsection{Continuum, 3.3$\,\umu$m band and the Pfund$\,\varepsilon$ line}
\label{sec:images}

Figure~1 shows the spatial distribution of \emph{L}-band continuum emission and the continuum-subtracted 3.3$\,\umu$m (total) feature.  The continuum emission is centrally peaked on HD~44179 and slightly extended along the bicone (`N~--~S') axis.
In contrast the overall 3.3$\,\umu$m band displays a pronounced bipolar structure.  It may be compared with that of \citet{mek98} for the inner \emph{c.} 2$''\times\,$2$''$, noting that in their study the data were obtained with a filter centered at 3.28$\,\umu$m with a spectral width of 0.16$\,\umu$m.  The image of \citet{mek98} differs significantly from the 3.3$\,\umu$m image in Figure~1 because it also contains continuum signal that was not subtracted and which was estimated to contribute \emph{c.} 60\% of the PAH emission signal.  Figure~1 shows that there is some 3.3$\,\umu$m PAH emission on-star (within the seeing constraint), with extended emission peaking \emph{c.} 0.7$''$ along both directions of the bipolar axis.  Notably the Southern maximum is stronger, as in the images of \citet{mek98} and \citet{men98}.  This has been interpreted as being due to inclination of the entire object which causes the Northern lobe emission to be partially obscured by dust in the circumbinary disc \citep{mek98}.

The results obtained from the double Lorentzian decomposition of the 3.3$\,\umu$m band are shown in Figure~2. These spatial distributions differ in the inner region with negligible 3.28$\,\umu$m~-~type emission directly towards HD~44179. Intensity profiles for the 3.3$\,\umu$m feature, and the `3.28'$\,\umu$m and `3.30'$\,\umu$m features are presented in Figure~3. The 3.28$\,\umu$m~-~type emission grows rapidly with increasing offset from the central star and peaks at the same positions (N and S) as for the 3.30$\,\umu$m emission, so apart from the innermost region the two images are quite similar.  It appears that a new form of PAH develops with increasing distance from the star, resulting in a set of PAHs that have C--H emission at a shorter wavelength.  The difference in the distributions seen in these IFU images provides evidence for two types of 3.3$\,\umu$m carrier, and supports the earlier conclusion from long-slit spectroscopic studies of the 3.3$\,\umu$m band profile in the Red Rectangle \citep{son03,son07}.

The image (Figure~4) for the continuum-subtracted feature near 3.4$\,\umu$m (see Table~1) is similar to that of the 3.28$\,\umu$m component and complements earlier spectroscopic observations by \citet{son03}, neither feature being present near to the star.  The 3.4$\,\umu$m emission is assigned to the C--H stretch arising through addition of a methyl group (CH$_3$) to the PAH edge or by hydrogenation as to make an sp$^3$~-~bonded carbon. Although not completely understood as discussed by \citet{tie08}, there is little doubt that the 3.4$\,\umu$m feature arises from a chemical addition to a standard aromatic PAH structure.  Figure~4 also shows the distribution of Pfund$\,\varepsilon$ hydrogen emission which traces the ionised central region of the nebula.

\begin{figure}
\includegraphics[scale=0.4]{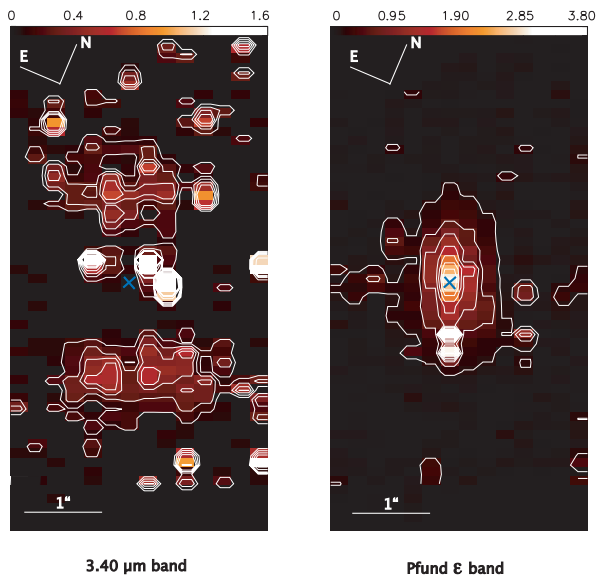}
\caption{Left panel. Spectral map of the 3.4~$\umu$m emission band.  Colour scale is in units of 10$^{-11}~$Wm$^{-2}\umu$m$^{-1}$. The contours correspond to 3, 6, 9, 12, 22, 32, 42 and 52\% of the maximum flux value. The 3.4~$\umu$m emission is not present in the vicinity of the central star and has a bipolar distribution. The Southern lobe appears stronger than the Northern equivalent. Right panel. Spectral map ($3.3''\times\,6.0''$) of the Pfund$\,\varepsilon$ hydrogen emission line.  The colour scale is in units of 10$^{-16}~$Wm$^{-2}\umu$m$^{-1}$.  The contours correspond to levels 5\% to 75\%, in steps of 10\%.}
\label{fig4}
\end{figure}

\section{Bay and non-bay character of small PAHs}

 Profile variation in the 3.3$\,\umu$m feature has been discussed by \citet{van04} with reference to charge state, size, edge structure and heteroatom substitution.  In particular these authors commented on differences that could arise according to whether the relevant hydrogen atoms occupied `non-bay' sites, as in very symmetrical compact PAHs, or `bay' sites where nearby C--H stretching motions could interfere; an example illustrating bay and non-bay sites is given in Figure~5 for the molecule perylene which has four bay and eight non-bay hydrogens.  The discussion of \citet{van04} was concerned principally with the influence of bay/non-bay hydrogens on overall line width rather than band frequency  and with specific reference to low-temperature matrix data which have multiple components. Based on the classifications of the 3.3$\,\umu$m and 11.2$\,\umu$m bands among different objects, it was concluded that molecular edge structure `is likely not the main cause of the variation in the FWHM of the 3.3$\,\umu$m feature' \citep{van04}.  We now examine `bay' and `non-bay' sites as a basis for explaining the 3.28 and 3.30$\,\umu$m spectral components with reference to laboratory and theoretical data.

\begin{figure}
\centering
\includegraphics[width=50mm]{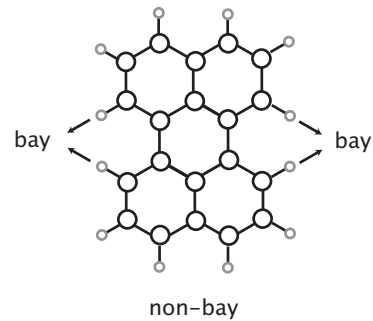}
\caption{Illustration of bay (four) and non-bay (eight) hydrogen sites in the molecule perylene, C$_{20}$H$_{12}$.}
\label{fig5}
\end{figure}

\subsection{Experimental data}
\subsubsection{3.3$\,\mu$m PAH emission following UV-pumping}
Infrared emission from PAHs in astrophysical environments is from vibrationally hot molecules at an effective temperature which depends mostly on the energy of the absorbed photon and molecular size. Laboratory infrared emission data obtained from from UV/vis pumped PAHs are optimal for comparison with astronomical spectra but the number of PAHs studied in this way is, unfortunately, small and the data cannot readily be used in evaluating the possible role of bay/non-bay character in relation to the 3.3$\,\umu$m band.

\subsubsection{3.3$\,\mu$m PAH absorption at low temperature}
Considering first experiments on PAHs in low-temperature inert gas matrices, the C--H stretching spectral region is known to be sensitive to the degree of bay/non-bay character, as remarked on by \citet{van04}.  Figure~1 of \citet{hud98a} and Figure~2 of \citet{hud98b}, show that molecules with non-bay~(nb) hydrogens only, such as pyrene with 10 non-bay hydrogens (C$_{16}$H$_{10}$; 10nb), have a single narrow dominant absorption feature with peak maximum at $\sim$~3050~cm$^{-1}$ in solid argon.  For a less symmetrical molecule with some bay~(b) hydrogens such as chrysene (C$_{18}$H$_{12}$; 8nb~+~4b), the spectral feature is broader with the C--H feature centred $\sim$20~cm$^{-1}$ to higher frequency. The triphenylene (6nb~+~6b) and hexabenzocoronene~A (C$_{42}$H$_{18}$; 6nb~+~12b) molecules have many more bay hydrogens and their strongest absorption is at around 3100~cm$^{-1}$ in the matrix environment. These spectral data indicate that the presence of `bay' sites introduces higher frequency absorption.  There also exist data for isolated jet-cold neutral gas-phase PAHs which, for the five PAHs in common, generally reflects the appearance of the matrix data with frequency agreement usually to within a few cm$^{-1}$ \citep{hun04}.

\subsubsection{3.3$\,\mu$m PAH absorption at high temperature}
We compare here 3.3$\,\umu$m observational data with infrared absorption spectra of heated gas-phase PAHs with internal temperatures of $\sim$500~K.  Data for a significant number of small to medium-sized PAHs are available from \citet{sem91}, \cite{job94,job95} and from the NIST Standard Reference Database \citet{ste11}. In Figure~6 infrared absorption spectra of four 4-ring PAH compounds are shown: the compact PAH pyrene (C$_{16}$H$_{10}$) with no bay hydrogens, and three 4-ring isomers (C$_{18}$H$_{12}$) -- benz[a]anthracene, chrysene and triphenylene which have 2, 4 and 6 bay hydrogens, respectively.  The spectroscopic data used in the figure are from the NIST Standard Reference Database \citep{ste11} and were recorded at $\sim$500~K. Similar recordings, with three molecules in common and one additional isomer, are given in Figure~7 of \citet{sem91}. The spectra for a given molecule are an average over many C--H stretch vibrational modes, including hot bands, and are remarkably simple in appearance. A particularly striking aspect of Figure~6 is the shift in band centre to shorter wavelength with increase in the number of bay hydrogens on the PAH structure.

\begin{figure}
\centering
\includegraphics[scale=0.15,width=60mm]{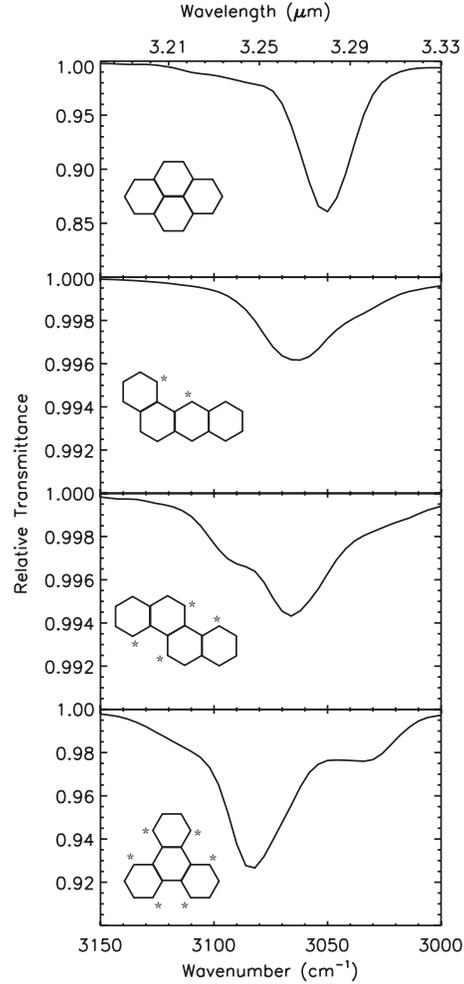}
\caption{Absorption spectra in the 3.3$\,\umu$m region from \citet{ste11} for (top to bottom) pyrene (10nb), benz[a]anthracene (10nb, 2b), chrysene (8nb, 4b) and triphenylene (6nb, 6b).  As per convention the hydrogens are not shown but those at bay sites are indicated by a star. }
\label{fig6}
\end{figure}

\begin{table}
\caption{Gas-phase absorption wavenumbers for PAHs from Table~2. of \citet{sem91} in which relative band strengths are given. The (nb, b) labelling and presentation of the wavenumbers in two columns is our work.  The molecules in rows 3-6 are the same as those in Figure~6.}
\begin{center}
\begin{tabular}{lccccc}
\hline
Molecule & Formula & nb & b & $\tilde\nu_1^a$ & $\tilde\nu_2^a$\\
%   &     &         &   cm$^{-1}$& $\,\mu$m   & Km mol$^{-1}$\\
\hline

Anthracene & C$_{14}$H$_{10}$ & 10 & 0 &  3063 & - \\
Phenanthrene & C$_{14}$H$_{10}$ & 8 & 2 &  3065 & - \\
\\
Pyrene & C$_{16}$H$_{10}$ & 10 & 0 &  3051$^b$& - \\
\\
Benz[a]anthracene & C$_{18}$H$_{12}$ & 10 & 2 &  3062& - \\
Chrysene& C$_{18}$H$_{12}$ & 8 & 4 &  3066 & 3089 \\
Triphenylene$^c$ & C$_{18}$H$_{12}$ & 6 & 6 & - & 3084 \\
Benzo[c]phenanthrene$^d$ & C$_{18}$H$_{12}$ & 10 & 2 &  3058& - \\
\\
Benzo[a]pyrene & C$_{20}$H$_{12}$ & 10 & 2 & 3056 & - \\
Benzo[e]pyrene & C$_{20}$H$_{12}$ & 8 & 4 & 3056 & 3086 \\
Perylene & C$_{20}$H$_{12}$ & 8 & 4 & 3059 & 3087 \\
\\
Dibenz[a,c]anthracene & C$_{22}$H$_{14}$ & 10 & 4 & 3066 & 3078 \\
Dibenz[a,h]anthracene & C$_{22}$H$_{14}$ & 10 & 4 & 3064 & 3077 \\

\hline
\label{semtable}
\end{tabular}
\end{center}
 {\bf Notes.} $a$. Wavenumber units. $b$. 3052~cm$^{-1}$ at 570~K \citep{job94}.
 $c$. A weaker band at 3032~cm$^{-1}$ is reported.  $d$. The interacting Hs have a different position (fjord) from the other cases in the Table; a weak band at 3013~cm$^{-1}$ is reported.
\end{table}

In their study of the absorption spectra of 33 gas-phase PAHs, \citet{sem91} reported the measurement of band wavenumbers, the most significant of which for this work are given in Table~2.  We have chosen to separate their reported wavenumbers into two columns labelled $\tilde\nu_1$ and $\tilde\nu_2$, which have mean values of 3061 and 3084~cm$^{-1}$ (3.267 and 3.243$\,\umu$m), respectively.  These values represent the summed contributions from many vibrational modes of each individual molecule as well as being averaged over the molecules listed in Table~2. They fall to the short wavelength side of the deduced `astronomical' components of 3.28 and 3.30$\,\umu$m \citep{son07}.  However, the experimental data of \citet{sem91} were recorded at $\sim$500~K which is a lower `temperature' than is expected for UV/vis pumping followed by IR cascade emission in an astronomical context for this size of molecule, and so a small shift to longer wavelength is expected. In the next Section a model taking into consideration the emission process of PAH molecules is presented.  A significant result is that the difference of 0.024$\,\umu$m in the average bay/non-bay laboratory data is close to the wavelength separation of $\sim$0.02$\,\umu$m obtained for the two components (3.28 and 3.30$\,\umu$m) from analysis of the astronomical data \citep{son03,son07}.  It therefore appears that the laboratory data fall largely into two frequency categories which, we suggest here, follow the degree of bay/non-bay character. This has much in common with the idea of an infrared `group frequency' where it is the chemical nature of a group that determines the vibrational frequency rather than it being specific to an individual molecule. Accordingly the lower (nb) and higher (b) C--H frequencies should hold for most PAH molecules.
Although the discussion in this Section has focussed on the bay/non-bay spectroscopic characteristics, it is important to note that the formation of bay-type sites arises naturally during growth of PAHs, for example in the growth from the symmetrical compact (non-bay) molecule pyrene into benzo[e]pyrene.  Hence PAH growth and the emergence of bay-type hydrogens are linked. In addition the similar spatial distributions of the 3.28$\,\umu$m and 3.40$\,\umu$m band carriers (Section~\ref{sec:images}) and as discussed in \citet{son07} suggests that the 3.28$\,\umu$m emission may also arise from addition of a chemical group (in this case C$_4$H$_2$) to the edge of a PAH structure which is consistent with formation of a `bay' configuration.

\section{Modelling of the C--H Band}
In order to test the bay/non-bay hypothesis, the photophysics of PAH molecules needs to be taken into account.  After the absorption of a UV photon, the energy of the molecule is redistributed in the vibrational modes through Intramolecular Vibrational Redistribution (IVR), and the molecule cools by emitting in its IR active modes (see for example \citealt{all89}). The approach used in modelling the 3.3$\,\umu$m feature consists of two parts: a) the vibrational infrared transition frequencies and transition strengths of a molecule are calculated using DFT and b) these data are used as input to compute the emission spectrum.

\subsection{DFT calculations}
 PAH vibrational transition frequencies and strengths were calculated using the B3LYP functional with a 6-31G* basis set within the Gaussian software package \citep{gauss}. This basis set was selected in preference to the more commonly used 4-31G basis as it provides better agreement between calculated and experimental intensities for the C--H stretching modes \citep{pat07}. Comparison of the computed DFT spectra with available experimental absorption spectra shows that the DFT method overestimates both the absolute C--H frequencies (as is well known) and also the frequency difference between bay and non-bay C--H stretches. The discrepancies are larger when B3LYP/4-31G is used. It is important to recognise that there is a significant redshift in the observed wavelength for high internal temperatures; for example at $\sim$500~K the peak wavelength (Table~2) for pyrene is 3.28$\,\umu$m whereas in emission following UV laser-desorption excitation the emission peak is at 3.30$\,\umu$m \citep{sch94} in good agreement with astrophysical observation.\\
To account for the differences between the DFT and experimental absolute and relative (\emph{i.e.} bay \emph{c.f.} non-bay) frequencies, two slightly different scaling factors were used and applied to all computed C--H stretching modes: 0.948 for the 2800~-~3210~cm$^{-1}$ range (non-bay) and 0.945 for $\omega>\,$3210~cm$^{-1}$ (bay).  The first factor (non-bay Hs) was determined by taking the experimental UV laser-desorbed experimental result of \citet{sch94} for pyrene with the band position-temperature law for pyrene as determined by \citet{job95}; a temperature of 1300~K for `astrophysical pyrene' is inferred from the observation, which corresponds to the absorption of a 6.35~eV photon in the UV-pumped modelling. The second scaling factor (for bay Hs) was calculated using the mean separation between experimental bay and non-bay modes from Table~\ref{semtable}. It is notable that the experimental bay \emph{c.f.} non-bay frequency difference does not change significantly with molecular size for the small-to-medium sized PAHs of relevance here. Inspection of Table~1 in \citet{bau09} for the C--H stretch modes in large PAH molecules (N$_C\,\geq\,$60) containing bay regions shows that the splitting with B3LYP/4-31G takes a range of higher values between 0.03 and 0.06$\,\umu$m but these large molecules are not thought to be the carriers of the 3.3$\,\umu$m band \citep{all89,sch93}.\\

\begin{figure}
\centering
\includegraphics[width=74mm]{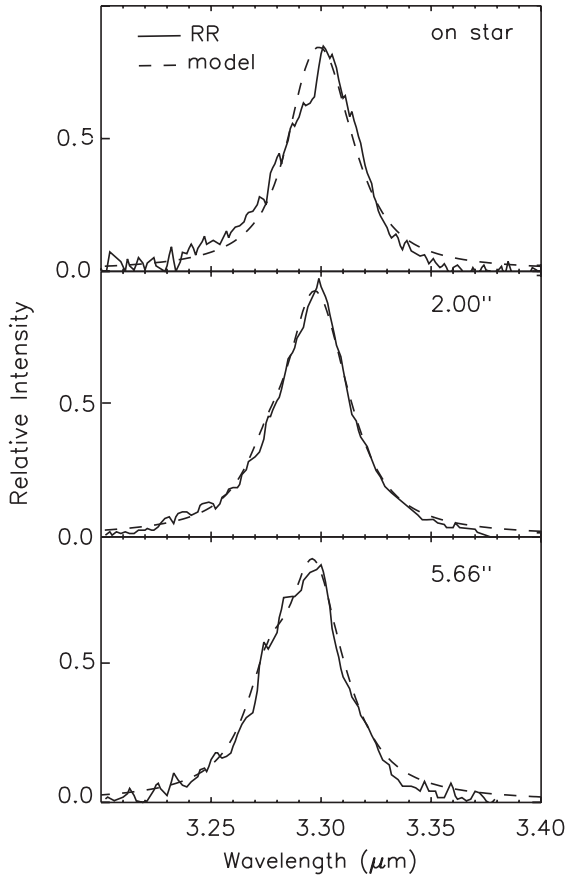}
\caption{Comparison between the 3.3$\,\umu$m band in the Red Rectangle at 0$''$, 2$''$ and 5.66$''$ offset from HD~44179 (solid line) and coadded emission spectra of pyrene and perylene (dashed line) in defined ratios (see text). Astronomical data are taken from \citet{son03} (top panel) and \citet{son07} (central and bottom panels). The Pf$\,\delta$ emission line at 3.296$\,\umu$m has been removed from the on-star spectrum, and is responsible for the slight deviation from a smooth profile at this wavelength. The band modelling is described in Section 6.}
\label{fig7}
\end{figure}

\subsection{The 3.3$\,\umu$m band profile}

The transition frequencies and strengths from the DFT calculations were used as input to a UV-pumped PAH emission model as employed by \citet{pat08}. In this model the rate of UV photon absorption by a PAH molecule is given by:$$R_{\mathrm{abs}}= \int_{3.2}^{13.6} \frac{B_{\nu} \sigma_{\nu}}{h\nu} d\nu$$  where $B_{\nu}$ is the Planck function of HD~44179 with $T_{\mathrm{eff}}=7750$~K \citep{men02}, $\nu$ is the UV photon frequency and the molecular photo-absorption cross-section, $\sigma_{\nu}$, is taken from the French-Italian database\footnote{http://astrochemistry.ca.astro.it/database/}\,\citep{mul06}. The emission spectrum arising from all bay and non-bay IR active C--H stretch transitions is then calculated. A Lorentzian with FWHM of 25~cm$^{-1}$ was used, this being the mean value deduced from the two-component spectroscopic fits of \cite{son07}. Anharmonic effects, for which there is little evidence in the form of `hot bands' in the astronomical spectra, and rotational broadening, were neglected.\\
Figure~\ref{fig7} shows a comparison between astronomical spectra of the Red Rectangle at three offsets and calculated emission spectra.  Pyrene was chosen as representative of a non-bay structure (10 nb; C$_{16}$H$_{10}$) and perylene (8 nb and 4 b; C$_{20}$H$_{12}$) as an example of a slightly larger molecule containing bay regions (Figure~\ref{fig5}).  In the top panel of Figure~\ref{fig7}, the on-star 3.30$\,\umu$m band is fitted well by a pyrene-only contribution. Indeed, \citet{son03} found that the on-star feature can be modelled by a single Lorentzian profile. The middle panel shows a coadded spectrum of pyrene and perylene (1:1.5 ratio) compared with the spectrum of the Red Rectangle at $2.00''$ offset along the NW axis. The presence of a molecule with bay regions (perylene), permits a very good fit of the 3.3$\,\umu$m band in a part of the nebula where the strength of the 3.28$\,\umu$m component starts to grow (\emph{c.f.} Figure~3 in \citealt{son07}). The bottom panel shows the astronomical spectrum at $5.66''$ offset compared with the coadded emission of pyrene and perylene in a 1:7 ratio (noting that perylene has 2:1 non-bay/bay hydrogens).  The increased intensity of the 3.28$\,\umu$m component further into the nebula can therefore be accounted for in terms of a more substantial contribution from bay-containing structures. This does not mean that perylene of itself gives rise to most of the $5.66''$ offset feature but rather that it would indicate an approximate ratio between non-bay and bay hydrogens of the order of 2:1 in the PAHs present at this offset. As the common Type~1 (A$_{3.3}$) feature is yet more symmetrical a more even ratio of non-bay/bay Hs is anticipated. This result is consistent with mass spectrometric analyses of PAHs in meteorites which reveal bay-containing PAHs to be common (see for example \citet{cal08} and references therein). Thus the bay/non-bay combination offers a satisfactory new explanation for the two components and profile variation in the 3.3$\,\umu$m band. It contrasts strongly with the proposal that the 3.3$\,\umu$m band is a superposition principally of anionic and neutral PAHs, the negatively charged PAHs being responsible for the longer wavelength part \citep{bau09}; this would require PAH anions to be the most abundant PAHs close to HD~44179 which seems unlikely.  Their ionisation energies are less than one eV and so these anions would be susceptible to electron photodetachment and hence removal of the carrier.
A larger set of small PAHs is required to test the bay/non-bay interpretation more fully and is beyond the scope of this paper.  However, there must be some question of uniqueness in the absence of additional information. It may also be noted that the transition strengths calculated using DFT cannot be fully relied on, perhaps particularly for subtle changes such as those between bay and non-bay environments; for example in the case of triphenylene (6b and 6nb) the experimental data (Figure~6) show the strongest absorption to fall in the the bay spectral region whereas the relatively high-level DFT calculations employed here predict bay and non-bay peaks to be of roughly equal intensity. Nevertheless, the fact that the observed 3.3$\,\umu$m profile and its evolution with offset can be accounted for based on experimental data, theoretical calculations and emission modelling leads to the conclusion that there is no need to invoke positively and/or negatively charged PAHs to explain the 3.3$\,\umu$m feature as discussed by \citet{bau08,bau09}.

\section{Other Red Rectangle emission}

The Red Rectangle is remarkable for its variety of spectral features in both absorption and emission across a range of wavelengths.  Unidentified optical emission bands and Extended Red Emission (ERE) appear prominently along the whiskers of the nebula.  However, the carriers of the relatively newly discovered `blue luminescence' are centred on HD~44179 \citep{vij04,vij05}.  This emission has been attributed to the fluorescence of optically pumped gas-phase PAHs including the small neutral molecules anthracene and pyrene. It might therefore be expected that the UIR emission from neutral PAHs and blue luminescence would share the same spatial distribution but this is not the case.  The blue luminescence is strongest near 3934~\AA\ and, while centrally peaked, it is clearly extended E~--~W along the direction of the circumbinary disk (see Figure~1 of \citealt{vij06}); this suggests a chemical or physical association between blue luminescence and the disk.  It is notable that the spatial distribution at 4050~\AA\ is, in contrast, highly symmetrical about the central star and is likely due predominantly to scattered starlight.

At $\sim$~5$''$S offset there is reasonable agreement between the 3.3$\,\umu$m and blue luminescence spatial distributions \citep{vij05}, but in the inner $\leq\,3''$ region they are markedly different.  It therefore appears that the carriers of blue luminescence and 3.3$\,\umu$m emission in the inner region are not the same entities.  This peculiarity could possibly be resolved if the blue luminescence were to arise from PAHs such as anthracene and pyrene adsorbed on cool grains in the disk ($\sim$~E--W), whereas the 3.3$\,\umu$m emission were due to gas-phase PAHs entrained in the $\sim$~N--S outflow. A recent near-IR polarimetric study of the inner part of the Red Rectangle has revealed scattering at \emph{J} which traces the circumbinary disk material and which appears also to follow the spatial distribution of blue luminescence \citep{gle09}. A measurement of the polarisation properties of blue luminescence would be of particular interest in assessing whether the blue luminescence carriers are free gas-phase molecules or molecules adsorbed on grains.

\section{Discussion and conclusions}

In earlier work \citep{son03,son07} it was proposed that the variation in profile of the 3.3$\,\umu$m PAH band with offset in the Red Rectangle could be interpreted in terms of the size distribution of neutral PAHs. Following the additional IFU observations reported here, and based on our analysis of additional laboratory data, consideration of the spectroscopic properties of bay/non-bay structures and emission modelling, we propose it is the emergence of bay-type structures of growing PAHs that underlies the observed profile change. The 3.3$\,\umu$m emission band of the Red Rectangle may then be interpreted in terms of compact non-bay PAHs dominating on-star with emission centred at 3.30$\,\umu$m, with larger PAH molecules which include bay-type structures (and a typical wavelength of 3.28$\,\umu$m) increasing in abundance with offset from HD~44179. The emergence of bay-type structures and the issue of PAH size are connected as the growth of PAHs must pass through stages where stoichiometric C$_4$H$_2$ entities are added to a PAH. Long slit spectra in the 10-15$\,\umu$m region in different positions of the nebula are needed to fully understand the variation of molecular edges, \emph{i.e.} bay and non-bay regions, and to compare with the different spatial distribution of the two components of the 3.3$\,\umu$m band.

We have considered other possible explanations for two 3.3$\,\umu$m emission band components including the effects of hydrogenation and methylation which are expected to increase with offset from the star HD~44179, and give rise to the 3.4$\,\umu$m emission band.  This band has a similar spatial distribution to that of the 3.28$\,\umu$m component - see Figure~2 and \citet{son03,son07}.  However, we have not found a systematic effect of either hydrogenation or methylation of PAH molecules on the 3.3$\,\umu$m aromatic C--H frequency in available laboratory and theoretical data \citep{wag00,pau99,pau01}.

Bay-type sites in a more general astronomical context can arise by addition of further rings as a PAH structure grows or alternatively through damage to the edge of a compact PAH. We favour the former interpretation for the inner part of the Red Rectangle given the good correlation in growth of the 3.28$\,\umu$m and 3.4$\,\umu$m features with offset, as the 3.4$\,\umu$m band is generally accepted as arising from addition of hydrogen and/or methyl groups.  In other astronomical environments such as H~{\sc ii} regions like Orion~H2S1 the observed very strong 3.28$\,\umu$m component \citep{son07} could arise because of damage to the edge structure. In the interpretation put forward here, the widely observed A$_{3.3}$ (Type~1) 3.3$\,\umu$m profile reflects a superposition of transitions of small neutral PAHs with bay (3.28$\,\umu$m) and non-bay (3.30$\,\umu$m) edges; it differs considerably from the conclusion of \citet{bau09} for the Red Rectangle where the 3.3$\,\umu$m feature is attributed to a mixture of anionic, neutral and (marginal) cationic PAHs.

In this paper we have described new IFU observations of the spatial distribution of PAHs that give rise to the 3.3$\,\umu$m emission band and a proposed two-component interpretation.  The images provide support for the proposal, based originally on spectroscopic long-slit data, that there are two types of C--H carrier that we suggest can be accounted for in terms of non-bay and bay hydrogen sites.

\section*{Acknowledgments}
We thank the UK Panel for the Allocation of Telescope Time for the award of observing time on UKIRT, STFC for a partial studentship and the University of Nottingham for studentship support to AC, the National Institute for International Education Department (NIIED) of the Korean Government
and The University of Nottingham for a studentship to In-Ok Song. We thank Amit Pathak for making the modelling code available, Keith Smith for a critical reading of the manuscript and an anonymous referee for useful comments which significantly improved this paper.

\label{lastpage}

\end{document}